\documentclass[11pt]{article}
\usepackage[table]{xcolor}
\usepackage{amsmath,amssymb,amsthm}
\usepackage[cp1250]{inputenc}
\usepackage{graphicx}
\usepackage{natbib}
\usepackage{setspace}
\usepackage{color}

\newtheorem{tw}{Theorem}%[section]

\newtheorem{asp}[tw]{Assumption}

\newtheorem{cor}[tw]{Corollary}
\newtheorem{prop}[tw]{Proposition}
\newcommand{\mtx}[1]{\boldsymbol{\mathrm{#1}}}
\newcommand{\vct}[1]{\lowercase{\boldsymbol{\mathrm{#1}}}}
\newcommand{\cmp}[2]{\lowercase{\mathrm{#1}}^{#2}}
\newcommand{\scl}[1]{\lowercase{\mathrm{#1}}}
\newcommand{\sclcal}[1]{\uppercase{\mathcal{#1}}}
\newcommand{\vctcal}[1]{\uppercase{\boldsymbol{\mathcal{#1}}}}
\newcommand{\lin}[1]{\mathcal{L}\left(#1\right)}
\newcommand{\normal}{\mathcal{N}\left(0,1\right)}
\newcommand{\E}[1]{\mathbb{E}\left[#1\right]}
\newcommand{\EP}[1]{\mathbb{E}^\mathbb{P}\left[#1\right]}
\newcommand{\EQ}[1]{\mathbb{E}^\mathbb{Q}\left[#1\right]}
\newcommand{\re}{\mathrm{e}}
\newcommand{\MGFQ}{\varphi^\mathbb{Q}_{\nu_1\nu_2}}
\newcommand{\MGFP}{\varphi^\mathbb{P}}

\numberwithin{equation}{section}

\begin{document}

\title{Smile from the Past:  A general option pricing framework \\ with multiple volatility and leverage components}
\author{Adam A. Majewski$^{\textrm{a,}}$\footnote{Corresponding author. Tel.: +39 05050 9094, \textit{E-mail address}: adam.majewski@sns.it}, Giacomo Bormetti$^{\textrm{a,b}}$, and Fulvio Corsi$^{\textrm{c,d}}$}
\date{April 2014}

\maketitle

\small
\begin{center}
  $^\textrm{a}$~\emph{Scuola Normale Superiore, Piazza dei Cavalieri 7, Pisa, 56126, Italy}\\
  $^\textrm{b}$~\emph{QUANTLab\hspace{2pt}\footnote{www.quantlab.it}, via Pietrasantina 123, Pisa, 56122, Italy}\\
  $^\textrm{c}$~\emph{Ca’ Foscari University of Venice, Fondamenta San Giobbe 873, Venezia, 30121, Italy}\\
  $^\textrm{d}$~\emph{City University London, Northampton Square, London EC1V 0HB, United Kingdom}
\end{center}
\normalsize

\smallskip

\begin{abstract} 
  In the current literature, the analytical tractability of discrete time option pricing models 
  is guaranteed only for rather specific types of models and pricing kernels.
  We propose a very general and fully analytical option pricing framework, encompassing a wide class of discrete time models
  featuring multiple-component structure in both volatility and leverage, and a flexible pricing kernel with multiple risk premia. 
  Although the proposed framework is general enough to include either GARCH-type volatility, Realized Volatility or a combination of the two, in this
  paper we focus on realized volatility option pricing models by extending the Heterogeneous Autoregressive Gamma (HARG) model
  of \citet{CorFusLav12} to incorporate heterogeneous leverage structures with multiple components, while preserving closed-form solutions for option prices. 
  Applying our analytically tractable asymmetric HARG model to a large sample of S\&P 500 index options, 
  we demonstrate its superior ability to price out-of-the-money options compared to existing benchmarks.
  \vspace{2cm}
\end{abstract}

%JEL:\\  
%Keywords: Realized volatility, Affine processes, Option pricing

\section{Introduction}
Due primarily to mathematical tractability and flexibility of incorporating 
various types of risk premia, the literature  on option pricing traditionally 
has been dominated by continuous-time processes.\footnote{ \cite{Heston93}, \cite{Duan95}, \cite{HN00},
 \cite{Merton76},  \cite{Bat96}, \cite{Bat00}, \cite{Pan02}, \cite{HuangWu04}, \cite{Bat06}, \cite{Eraker04}, \cite{EJP03} and \cite{BCJ07} }
On the other hand, models for asset dynamics under the physical measure $\mathbb{P}$ have primarily been developed in discrete-time.
The time-varying volatility models of the ARCH-GARCH families 
\citep{Engle82, Bol86, GJR93, Nelson91} have led the field in estimating and predicting the volatility dynamics. 
More recently, thanks to the availability of intra-day data, the so 
called Realized Volatility (RV) approach also became a prominent approach for measuring 
and forecasting volatility. 
The key advantage of the RV is that it provides a precise nonparametric measure
of daily volatility\footnote{This idea trace back to \cite{MRT.1980} and has been recently formalized and generalized in a series
of papers that apply the quadratic variation theory to the 
class of $L_2$ semi-martingales; See, e.g., \cite{Comte98}, \cite{ABDL01}
\cite{ABDL03}, \cite{BS2001}, \cite{BS2002a}, \cite{BS2002b}, \cite{BS2002c}.} 
(i.e., making it observable) which leads to simplicity in model estimation and superior forecasting performance.\\

Discrete time models present the important advantage of being easily filtered and estimated even in the presence of complex dynamical features 
such as long memory, multiple components and asymmetric effects, which turns out to be crucial in improving volatility forecast and option pricing performances.
A growing strand of literature advocates for the presence of a multi-factor volatility structure both under the physical measure \citep{Muletal97, EngLee99, BolWri01, BS2001, CalvetFisher04} 
and the risk neutral one \citep{Bat00, Bat12, LiZhang10, ChrJacOrnWan08, AdrRos07}.
In the discrete time option pricing literature, multiple components have
been incorporated into both GARCH-type \citep{ChrJacOrnWan08} and realized volatility models \citep{CorFusLav12}, and both approaches
have shown that short-run and long-run components are necessary to capture the term structure of
the implied volatility surface. 
Also in the modelling of the so called leverage effect (the asymmetric impact of positive and negative past returns on future volatility), 
recent papers advocate the need for a multi-component leverage structure in volatility forecasting~\citep{SchMed09, CorRen12}.
Finally, the need for a flexible pricing kernel incorporating variance-dependent risk premia, in addition to the common equity risk premium, 
has been well-documented by \citet{CHJ11}.
However, in the current literature, the analytical tractability of discrete time option pricing models 
is guaranteed only for rather specific types of models and pricing kernels.\\

The purpose of this paper is to propose a very general framework encompassing a wide class of discrete time multi-factor asymmetric volatility models
for which we show how to derive (using conditional moment-generating functions) closed-form option valuation
formulas under a very general and flexible state-dependent pricing kernel.
This general framework allows for a wide range of interesting applications. 
For instance, it permits a straightforward generalization of both the multi-component GARCH-type model of \citet{ChrJacOrnWan08}
as well as of the Heterogeneous Autoregressive Gamma (HARG) model for realized volatility of \citet{CorFusLav12}.
In this paper we focus our attention on the applications of the general framework to the realized volatility class of model, while its applications
to the GARCH type of model will be the subject of a separate, companion paper.\\

In more detail, this paper provides several theoretical results for both the general framework and for the 
specific application to realized volatility models which can be summarized as follows.
For the general framework we show: 
(i) the recursive formula for the analytical Moment Generating Function (MGF) under $\mathbb{P}$,
(ii) the general characterization of the analytical no-arbitrage conditions,
(iii) the formal change of measure obtained using a general and flexible exponentially affine Stochastic Discount Factor (SDF), 
which features both equity risk premium and multi-factor variance risk premia, 
(iv) the recursive formula for the analytical MGF under $\mathbb{Q}$.\\

In addition, by applying the general framework to the specific class of model featuring HARG type dynamics for realized volatility we are able to:
(i) introduce various flexible types of leverage with heterogeneous structures analogous to the one specified by the HARG model for volatility, 
by preserving the full analytical tractability of the model,
(ii)  have flexible skewness and kurtosis term structure under both $\mathbb{P}$ and $\mathbb{Q}$,
(iii) have an explicit one-to-one mapping between the parameters of the volatility dynamics under $\mathbb{P}$ and $\mathbb{Q}$,
(iv)  have closed-form option prices for  model with heterogeneous realized volatility and leverage dynamics.
Finally, by applying our fully analytically tractable HARG model with heterogeneous leverage
on a large sample of S\&P 500 index options, we show the superior ability of the model in pricing out-of-the-money (OTM) options
compared to existing benchmarks.\\

The rest of the paper is organized as follows. In Section~\ref{sec:MFV} we propose the general framework for option pricing with multi-factor volatility models. Section~\ref{sec:LHARG-RV} defines a family of HARG models for realized volatility with leverage (LHARG), presents two particular models belonging to the family, describes the estimation of the models, and analyzes their statistical properties. Section~\ref{sec:Valuation} reports the option pricing performance of LHARG models, comparing them to benchmark models. Finally, in Section~\ref{sec:Conclusions} we summarize the results.

\section{The multi-factor volatility models}
\label{sec:MFV}

\subsection{General framework}
The main purpose of introducing a multi-factor structure in volatility modeling is to account for dependencies among volatilities at different time-scales. Currently, there are two alternative approaches in the literature. The first is to decompose the daily volatility into several factors and model the dynamics of each factor independently, as done by~\cite{ChrJacOrnWan08} or~\cite{FouLor11} in terms of short-run and long-run volatility components. The other approach is to define factors as an average of past volatilities over different time horizons, for instance the daily, weekly and monthly components in~\cite{Cor09}. In this section we propose a general framework which includes both approaches.\\

We consider a risky asset with price $S_t$ and geometric return
\begin{equation*}
  y_{t+1} =  \log \left( \frac{S_{t+1}}{S_t} \right).
\end{equation*}
To model the dynamics of log-returns we define the $k$-dimensional vector of factors $\cmp{F}{1}_{t}$, $\ldots$, $\cmp{F}{k}_{t}$ which we shortly denote as $\vct{F}_t$. The volatility on day $t$ is defined as a linear function of factors $\mathcal{L}:\mathbb{R}^k \rightarrow \mathbb{R}$ and the daily log-returns on day $t+1$ are modeled by equation 
\begin{equation}
  \label{logreturns}
  y_{t+1} =  r +  \lambda ~\lin{\vct{F}_{t+1}}  + \sqrt{\lin{\vct{F}_{t+1}}} ~\epsilon_{t+1}\,,
\end{equation}
where $r$ is the risk-free rate, $\lambda$ is the market price of risk, and $\epsilon_{t}$ are i.i.d. $\normal$. We model $\vct{F}_{t+1}$ as 
\begin{equation}
  \label{dist}
  \vct{F}_{t+1} | \mtx{F}_t, \mtx{L}_t \sim \mathcal{D} \left(\Theta_0,\Theta(\mtx{F}_t,\mtx{L}_t)\right)\,,
\end{equation}
where $\mathcal{D}$ denotes a generic distribution depending on the vector of parameters $\Theta$ which is a $k$-dimensional function of the matrices $\mtx{F}_t = (\vct{F}_t,\ldots,\vct{F}_{t-p+1}) \in \mathbb{R}^{k\times p}$ and $\mtx{L}_t = (\vct{\ell}_t,\ldots,\vct{\ell}_{t-q+1}) \in \mathbb{R}^{k\times q}$ for $p>0$ and $q>0$, respectively. 
We consider the case of a linear dependence of $\Theta$ on $\mtx{F}$ and $\mtx{L}$
\begin{equation}
  \label{framework}
  \Theta(\mtx{F}_t,\mtx{L}_t) = \vct{d} + \sum_{i=1}^{p} \mtx{M}_i \vct{F}_{t+1-i} + \sum_{j=1}^{q} \mtx{N}_j \vct{\ell}_{t+1-j}\,, 
\end{equation}
where $\mtx{M}_i, \mtx{N}_j \in \mathbb{R}^{k\times k}$ for $i=1,\ldots,p$ and $j=1,\ldots,q$, $\vct{d} \in \mathbb{R}^{k}$, and vectors $\vct{\ell}_{t-j}$ are of the form 
\begin{equation}
  \label{lev}
  \vct{\ell}_{t+1-j} = \begin{bmatrix} \left(\epsilon_{t+1-j} - \gamma_1 \sqrt{\lin{\vct{F}_{t+1-j}}}\right)^2 \\ \vdots \\ \left(\epsilon_{t+1-j} - \gamma_k  \sqrt{\lin{\vct{F}_{t+1-j}}}\right)^2  \end{bmatrix}. 
\end{equation}
The vector $\Theta_0$ collects all the parameters of the distribution $\mathcal{D}$ which do not depend on the past history of the factors and of the leverage. For the distribution $\mathcal{D}$ considered in this paper (Dirac delta and non-central Gamma distribution) the sufficient condition for the non-negativity of process reads:
\begin{equation}
\label{eq:poscon}
 \vct{d} \geq 0 \ \ \ \ \mtx{M}_i  \geq 0 \ \ \mbox{for all } i \in \{1 , \ldots, p \} \ \ \ \   \mtx{N}_j \geq 0  \ \ \mbox{for all } j \in \{1 , \ldots, q \},
\end{equation}
where $\geq$ has to be meant as componentwise inequality.\\

The results presented in this paper are derived under the general assumption
\begin{asp}
\label{assumption}
 The following relation holds true 
  \begin{equation}
    \label{affine}
    \E{\re^{ z y_{s+1} + \vct{b}\cdot\vct{F}_{s+1} + \vct{c}\cdot\vct{\ell}_{s+1} }| \mathcal{F}_{s}} = \re^{\sclcal{A}(z,\vct{b},\vct{c}) + \sum_{i=1}^{p}\vctcal{B}_i(z,\vct{b},\vct{c})\cdot\vct{F}_{s+1-i} + \sum_{j=1}^{q}\vctcal{C}_j(z,\vct{b},\vct{c})\cdot\vct{\ell}_{s+1-j} }
  \end{equation}
  for some functions $\sclcal{A}:\mathbb{R}\times\mathbb{R}^k\times\mathbb{R}^k \rightarrow \mathbb{R}$, $\vctcal{B}_i:\mathbb{R}\times\mathbb{R}^k\times\mathbb{R}^k \rightarrow \mathbb{R}^k$, and $\vctcal{C}_j:\mathbb{R}\times\mathbb{R}^k\times\mathbb{R}^k \rightarrow \mathbb{R}^k$, where $\vct{b},\vct{c} \in \mathbb{R}^k$ and $\cdot$ stands for the scalar product in $\mathbb{R}^k$.  
\end{asp}

Our framework is suited to include both GARCH-like models and realized volatility models. As far as the former class is concerned, we encompass the family of multiple component GARCH models with parabolic leverage pioneered in~\cite{HN00} and later extended to the two Component GARCH (CGARCH) by~\cite{ChrJacOrnWan08}. For instance, the latter model corresponds to the following dynamics
\begin{equation}
  \label{cgarch}
  \begin{split}
    y_{t+1} &= r +  \lambda h_{t+1}  + \sqrt{h_{t+1}} \epsilon_{t+1}\,,\\
    h_{t+1} &= q_{t+1} + \beta_1 \left( h_{t} - q_{t} \right)  + \alpha_1 \left(\epsilon_t^2 -1 -2\gamma_1 \epsilon_t \sqrt{h_t }\right)\,,\\
    q_{t+1} &= \omega + \beta_2 q_{t} + \alpha_2 \left(\epsilon_t^2 -1 -2\gamma_2 \epsilon_t \sqrt{h_t }\right)\,.
  \end{split}
\end{equation}
Setting $k=2$, we define $\cmp{F}{1}_{t+1} = h_{t+1} - q_{t+1}$ and $\cmp{F}{2}_{t+1} = q_{t+1}$ and rewrite the model as 
\begin{equation}
  \label{eq:CJOWmap}
  \begin{bmatrix} 
    \cmp{F}{1}_{t+1}  \\ 
    \cmp{F}{2}_{t+1}  
  \end{bmatrix} 
  = 
  \begin{bmatrix}
    - \alpha_1 \\ 
    \omega - \alpha_2 
  \end{bmatrix}
  +
  \begin{bmatrix} 
    \beta_1 -\alpha_1 \gamma_1^2 & -\alpha_1 \gamma_1^2  \\ 
    -\alpha_2 \gamma_2^2 & \beta_2 -\alpha_2 \gamma_2^2 
  \end{bmatrix} 
  \begin{bmatrix} 
    \cmp{F}{1}_t  \\ 
    \cmp{F}{2}_t  
  \end{bmatrix} 
  +
  \begin{bmatrix} 
    \alpha_1 & 0 \\ 
    0 &  \alpha_2   
  \end{bmatrix} 
  \begin{bmatrix} 
    \left(\epsilon_t - \gamma_1 \sqrt{\lin{\vct{F}_t}}\right)^2 \\ 
    \left(\epsilon_t - \gamma_2 \sqrt{\lin{\vct{F}_t}}\right)^2 
  \end{bmatrix}\,, 
\end{equation}
where $\lin{\vct{F}_t}=\cmp{F}{1}_t+\cmp{F}{2}_t=h_t$. If we now specify for $\mathcal{D}$ in eq.~(\ref{dist}) the form of a Dirac delta distribution, define $\vct{d}=\left(-\alpha_1,~\omega-\alpha_2\right)^t$, and identify the matrices $\mtx{M}_1$ and $\mtx{N}_1$ in a natural way from the right term side of eq.~(\ref{eq:CJOWmap}), the model by Christoffersen \textit{et al.} fits the general formula~(\ref{dist}). It is worth mentioning that for the CGARCH model it is not possible to ensure the non-negative definiteness of both $h_{t}$ and $q_{t}$ for all $t$ (condition (\ref{eq:poscon}) is not satisfied). Nonetheless, for realistic values of the parameters the probability of obtaining negative volatility factors is extremely low, and this drawback is largely compensated for by the effectiveness of the model in capturing real time series empirical features. We discuss the issue of positivity in greater detail in Section~\ref{sec:ESTIMATION-STAT}.\\

The second example that we discuss is the class of realized volatility models known as Autoregressive Gamma Processes (ARG) introduced in~\cite{GouJas06}, to whom the Heterogeneous Autoregressive Gamma (HARG) model presented in~\cite{CorFusLav12} belongs. The process $\mathrm{RV}_t$ is an ARG(p) if and only if its conditional distribution given $\left(\mathrm{RV}_{t-1},\ldots,\mathrm{RV}_{t-p}\right)$ is a noncentred gamma distribution $\bar\gamma(\delta,\sum_{i=1}^p \beta_i \mathrm{RV}_{t-i},\theta)$, where $\delta$ is the shape, $\sum_{i=1}^p \beta_i \mathrm{RV}_{t-i}$ the non-centrality, and $\theta$ the scale. Then, the model described by eq.s~(\ref{dist})-(\ref{framework}) reduces to an ARG(p) if we fix $k=1$, $\cmp{F}{}_t=\mathrm{RV}_t$, $\mathcal{D}=\bar{\gamma}\left(\Theta_0,\Theta(\mtx{F}_{t-1})\right)$ with
\begin{equation*}
  \Theta_0 = \left(\delta, \theta\right)^t\,,\quad\mbox{and}\quad\Theta(\mtx{F}_{t-1})=\sum_{i=1}^p \beta_i \cmp{F}{}_{t-i}\,.
\end{equation*}

\subsection{Physical and risk-neutral worlds}
\label{sec:PRNW}
The general framework defined by eq.s~(\ref{logreturns})-(\ref{lev}) combined with the assumption~(\ref{affine}) allows us to completely characterize the MGF of the log-returns under the physical measure. If relation (\ref{affine}) is satisfied, then the moment generating function of $\ln(S_T/S_t)$ is given by recursive relation in terms of functions $\sclcal{A}$, $\sclcal{B}_i$, $\sclcal{C}_j$: we present the formulae in Appendix~\ref{app:MGFCOMPUTATION}.\\

By specifying the Stochastic Discount Factor (SDF) within the family of the exponential-affine factors, we are able to compute 
analogous recursions under $\mathbb{Q}$. 
The need for variance-dependent risk premia in SDF, in addition to the common equity risk premium, 
has been shown by \citet{CHJ11}, \citet{GagGouRen11} and \citet{CorFusLav12} to be crucial in reconciling the time series properties of stock returns with the cross-section of option prices. 
Our framework permits the adoption of a very general and flexible pricing kernel incorporating, in addition to the common equity risk premium, 
multiple factor-dependent risk premia.
The most general SDF that we might consider in our framework corresponds to the following
\begin{equation}
  \label{SDF}
  M_{s,s+1} = \frac{\re^{-\vct{\nu}\cdot\vct{F}_{s+1} - \nu_{2} y_{s+1} }}{\EP{\re^{-\vct{\nu}\cdot\vct{F}_{s+1} - \nu_{2} y_{s+1}} | \mathcal{F}_{s}}}\,,
\end{equation}
with $\vct{\nu}\in\mathbb{R}^k$. 
The general framework allows us to introduce $k+1$ risk premia. In this paper we consider models where $\vct{F}_t$ is one-dimensional and corresponds to the continuous component of the realized variance. Thus we restrict to two risk premia, $\nu_1$ and $\nu_2$, leaving open the possibility for future research to include further risk premia related to other volatility components due to jumps and overnight returns.\\
 
Moment generating function under risk-neutral measure for models where the joint dynamics of log-returns and volatiltiy is affine combined with exponential-affine SDF, can be derived in semi-closed form, as has been shown in \cite{GouMon07}. In Appendix \ref{app:MGFCOMPUTATION} we show that MGF of $\ln(S_T/S_t)$ under risk-neutral measure $\mathbb{Q}$ is given by recursive relation in terms of functions $\sclcal{A}$, $\sclcal{B}_i$, $\sclcal{C}_j$. The resulting risk-neutral dynamics depend on the values of the equity and variance risk premia, $\nu_1$ and $\nu_2$ respectively, which have to satisfy the no arbitrage constraints. For all the models within the general framework, the no-arbitrage conditions can be written in terms of functions $\sclcal{A}$, $\sclcal{B}_i$, $\sclcal{C}_j$ given in Propositon \ref{th:NOARBITRAGE} which summarizes all the results of this section.
\begin{prop}
\label{th:NOARBITRAGE}
If Assumption \ref{assumption} is satisfied then the moment generating function of $\ln(S_T/S_t)$ under measures $\mathbb{P}$ and $\mathbb{Q}$ is given by recursive relation in terms of functions $\sclcal{A}$, $\sclcal{B}_i$, $\sclcal{C}_j$. Moreover the SDF~(\ref{SDF}) is compatible with the no arbitrage restriction if the following conditions are satisfied:
\begin{equation}
  \label{eq:noARB}
  \begin{split}
    \sclcal{A}    (1- \nu_{2},-\vct{\nu}_{1},\vct{0}) &= r + \sclcal{A}(- \nu_{2},-\vct{\nu}_{1},\vct{0})\\
    \vctcal{B}_{i}(1- \nu_{2},-\vct{\nu}_{1},\vct{0}) &= \vctcal{B}_{i}(- \nu_{2},-\vct{\nu}_{1},\vct{0}) \ \ \text{for} \ i=1,\ldots,p\\
    \vctcal{C}_{j}(1- \nu_{2},-\vct{\nu}_{1},\vct{0}) &= \vctcal{C}_{j}(- \nu_{2},-\vct{\nu}_{1},\vct{0}) \ \ \text{for} \ j=1,\ldots,q.
  \end{split}
\end{equation}
\end{prop}
Proof: See Appendix \ref{app:NOARBITRAGE} and \ref{app:MGFCOMPUTATION} .\

\section{LHARG-RV}
\label{sec:LHARG-RV}
\subsection{The model}
\label{sec:THEMODEL}

HAR-RV processes were introduced to financial literature by~\cite{Cor09}, and are characterized by the different impact that past realized variances aggregated on a daily, weekly and monthly basis have on today's realized variance. Lagged terms are collected in three different non-overlapping factors: $\mathrm{RV}_t$ (short-term volatility factor), $\mathrm{RV}_t^{(w)}$ (medium-term volatility factor), and $\mathrm{RV}_t^{(m)}$ (long-term volatility factor). \cite{CorFusLav12} presents the application of HAR-RV models to option pricing, discussing an extension of the HAR-RV which includes a daily binary Leverage component (HARGL). However, in~\cite{CorRen12} the authors stress the importance of a heterogeneous structure for leverage. Thus we develop an Autoregressive Gamma model with Heterogeneous parabolic Leverage, and we name it the LHARG-RV model.\\

LHARG-RV belongs to the family of models described by~(\ref{logreturns})-(\ref{lev}) setting $k=1$ and $\cmp{F}{}_t = \mathrm{RV}_t$. Thus, log-returns evolve according to the equation
\begin{equation}
  \label{eq:lrdyn}
  y_{t+1} =  r +  \lambda \mathrm{RV}_{t+1}  + \sqrt{\mathrm{RV}_{t+1}} \epsilon_{t+1}\,,
\end{equation}
while the realized variance at time $t+1$ conditioned on information at day $t$ is sampled from a noncentred gamma distribution
\begin{equation}
  \label{RVdyn}
  \mathrm{RV}_{t+1} | \mathcal{F}_{t} \sim \bar{\gamma} (\delta, \Theta(\mtx{RV}_t,\mtx{L}_t), \theta)
\end{equation}
with
\begin{equation}
  \label{thetalharg}
  \Theta(\mtx{RV}_t,\mtx{L}_t)  = d + \beta_d \mathrm{RV}_t^{(d)} + \beta_w \mathrm{RV}_t^{(w)} + \beta_m \mathrm{RV}_t^{(m)}
  + \alpha_d \scl{\ell}_t^{(d)} + \alpha_w \scl{\ell}_t^{(w)} + \alpha_m \scl{\ell}_t^{(m)}\,. 
\end{equation}
In the previous equation $d\in \mathbb{R}$ is a constant and the quantities
\begin{equation*}
  \begin{array}{ll}
    \mathrm{RV}_t^{(d)} = \mathrm{RV}_t,  & \scl{\ell}_t^{(d)} =  \left( \epsilon_{t} - \gamma \sqrt{\mathrm{RV}_{t}}\right)^2,\\
    \mathrm{RV}_t^{(w)} = \frac{1}{4}\sum_{i=1}^{4}\mathrm{RV}_{t-i}, &  \scl{\ell}_t^{(w)} = \frac{1}{4}\sum_{i=1}^{4}\left( \epsilon_{t-i} -  \gamma \sqrt{\mathrm{RV}_{t-i}}\right)^2,\\
    \mathrm{RV}_t^{(m)} = \frac{1}{17}\sum_{i=5}^{21}\mathrm{RV}_{t-i}, & \scl{\ell}_t^{(m)} = \frac{1}{17}\sum_{i=5}^{21}\left( \epsilon_{t-i} - \gamma \sqrt{\mathrm{RV}_{t-i}}\right)^2,
  \end{array}
\end{equation*}
correspond to the heterogeneous components associated with the short-term (daily), medium-term (weekly), and long-term (monthly) volatility and leverage factors, on the left and right columns respectively. The structure of leverage is analogous to the one in \cite{HN00}, and it is based on asymmetric influence of shock: large positive idiosyncratic component $\epsilon_t$ has a smaller impact on $RV_{t+1}$ than large negative $\epsilon_t$. As consequence the log-returns and variance process are negatively correlated:
\begin{equation}
\begin{split}
Cov_{t-1} (y_t , RV_{t+1})  &= - 2 \theta \alpha_d  \gamma \E{ RV_t    | \mathcal{F}_{t-1} }\\
&= -2\theta^2 \alpha_d \gamma \left(   \delta +  \Theta(\mtx{RV}_{t-1},\mtx{L}_{t-1}) \right)\,. 
\end{split}
\end{equation} 
In order to adjust eq.~(\ref{thetalharg}) to our framework we rewrite $\Theta(\mtx{RV}_t,\mtx{L}_t)$ as
\begin{equation}
  \label{tilderv}
  d + \sum_{i=1}^{22} \beta_i \mathrm{RV}_{t+1-i} + \sum_{j=1}^{22} \alpha_j \left(\epsilon_{t+1-j} - \gamma \sqrt{\mathrm{RV}_{t+1-j}}\right)^2\,,
\end{equation}
with
\begin{equation}
  \label{alphanbeta}
  \beta_i = 
  \left\{ 
    \begin{array}{ll} 
      \beta_d      &\mbox{ for  } i = 1 \\  
      \beta_w/4    &\mbox{ for  } 2 \leq i \leq 5 \\  
      \beta_m/17   &\mbox{ for  } 6 \leq i \leq 22   
    \end{array} 
    \right.  
  \quad 
  \alpha_j = 
  \left\{ 
    \begin{array}{ll} 
      \alpha_d      &\mbox{ for  } j = 1 \\  
      \alpha_w/4    &\mbox{ for  } 2 \leq j \leq 5 \\  
      \alpha_m/17   &\mbox{ for  } 6 \leq j \leq 22   
    \end{array} 
  \right.\,.
\end{equation}

We show in Appendix~\ref{app:MGFCOMPUTATION} that LHARG models satisfy Assumption~\ref{assumption}, and we explicitly derive the $\sclcal{A}$, $\sclcal{B}_i$, and $\sclcal{C}_j$ functions. Then, the general results presented in Section~\ref{sec:PRNW} read

\begin{prop}
  \label{prop:mgfP}
  Under $\mathbb{P}$, the MGF for LHARG model has the following form
  \begin{equation}
    \label{Pmgf}
    \MGFP(t,T,z) = \EP{\re^{z y_{t,T} } | \mathcal{F}_{t}} = \exp \left( \scl{A}_{t} + \sum_{i=1}^{p} \scl{B}_{t,i} \mathrm{RV}_{t+1-i} + \sum_{j=1}^{q} \scl{C}_{t,j} \scl{\ell}_{t+1-j}  \right) 
  \end{equation}
  where 
  \begin{equation}
    \label{coefmgfP}
    \begin{split}
      \scl{A}_{s} &= \scl{A}_{s+1} + zr - \frac{1}{2} \ln(1-2\scl{C}_{s+1,1}) - \delta \sclcal{w}( \scl{X}_{s+1} ,\theta) + d \sclcal{v}(\scl{X}_{s+1},\theta)\\
      \scl{B}_{s,i} &=
      \begin{cases} 
	\scl{B}_{s+1,i+1} +  \sclcal{v}(\scl{X}_{s+1} ,\theta)  \beta_i  &\mbox{for} \ 1 \leq i \leq p-1 \\ 
	\sclcal{v}(\scl{X}_{s+1} ,\theta)   \beta_i  &\mbox{for} \ i = p
      \end{cases}\\
      \scl{C}_{s,j} &= 
      \begin{cases} 
	\scl{C}_{s+1,j+1} + \sclcal{v}(\scl{X}_{s+1} ,\theta)  \alpha_j  &\mbox{for} \ 1 \leq j \leq q-1 \\  
	\sclcal{v}(\scl{X}_{s+1} ,\theta) \alpha_j  &\mbox{for} \ j = q 
      \end{cases}
    \end{split}
  \end{equation}
  with 
  \begin{equation*}
    \scl{X}_{s+1} = z\lambda + \scl{B}_{s+1,1} + \frac{\frac{1}{2}z^2 + \gamma^2\scl{C}_{s+1,1} - 2\scl{C}_{s+1,1} \gamma z}{1-2\scl{C}_{s+1,1}}\,.
  \end{equation*}
  The functions $\sclcal{v}$, $\sclcal{w}$ are defined as follows
  \begin{equation}
    \sclcal{v}(x,\theta)=\frac{\theta x}{1 - \theta x} \ \ \ \ \mbox{  and  } \ \ \ \ \sclcal{w}(x,\theta)= \ln \left( 1 - x \theta \right)\,,
  \end{equation}
  and the terminal conditions read $\scl{A}_T = \scl{B}_{T,i} = \scl{C}_{T,j} = 0$ for $i=1,\ldots,p$ and $j=1,\ldots,q$.
\end{prop}
Proof: See Appendix \ref{app:LHARG}.\\

The proof of the previous proposition provides us with the explicit form of the functions $\sclcal{A}$, $\sclcal{B}_i$, and $\sclcal{C}_j$ for the general class of LHARG models. Following the reasoning in Appendix F in \cite{GouJas06} one can derive the stationarity condition for $RV_t$ process:
\begin{equation}
\theta \left( \beta_d + \beta_w + \beta_m + \gamma^2 \left(  \alpha_d + \alpha_w + \alpha_m \right)  \right) < 1.
\end{equation}
Employing the SDF suggested in~(\ref{SDF}), which for LHARG takes the form
\begin{equation}
  \label{LHARGSDF}
  M_{s,s+1} = \frac{\re^{-\nu_{1}  \mathrm{RV}_{s+1} - \nu_{2} y_{s+1} }}{\EP{\re^{-\nu_{1} \mathrm{RV}_{s+1}  - \nu_{2} y_{s+1}} | \mathcal{F}_{s}}}\,,
\end{equation}
and plugging the $\sclcal{v}$ and $\sclcal{w}$ functions in eq.~(\ref{eq:GF-MGFQrecursive}) we readily obtain the risk-neutral MGF. 

\begin{cor}
Under the risk-neutral measure $\mathbb{Q}$ the MGF for LHARG has the form
\begin{equation*}
  \MGFQ(t,T,z) = \exp \left( \scl{A}^*_{t} + \sum_{i=1}^{p} \scl{B}^*_{t,i} \mathrm{RV}_{t+1-i} + \sum_{j=1}^{q} \scl{C}^*_{t,j} \scl{\ell}_{t+1-j}\right)\,, 
\end{equation*}
where 
\begin{equation}
  \label{coefmgfQ}
  \begin{split}
    \scl{A}^*_{s} =& \scl{A}^*_{s+1} + zr - \frac{1}{2} \ln(1-2\scl{C}^*_{s+1,1}) - \delta \sclcal{w}( \scl{X}^*_{s+1} ,\theta) + \delta \sclcal{w}( \scl{Y}^*_{s+1} ,\theta) \\
             &+  d \sclcal{v}( \scl{X}^*_{s+1} ,\theta) - d \sclcal{v}( \scl{Y}^*_{s+1},\theta)\\
    \scl{B}^*_{s,i} =&
    \begin{cases} 
      \scl{B}^*_{s+1,i+1} + \left( \sclcal{v}(\scl{X}^*_{s+1} ,\theta) - \sclcal{v}(\scl{Y}^*_{s+1} ,\theta) \right) \beta_i  &\mbox{for} \ 1 \leq i \leq p-1 \\ 
      \left( \sclcal{v}(\scl{X}^*_{s+1} ,\theta) - \sclcal{v}(\scl{Y}^*_{s+1} ,\theta) \right) \beta_i  &\mbox{for} \ i = p
    \end{cases}\\
    \scl{C}^*_{s,i} =& 
    \begin{cases} 
      \scl{C}^*_{s+1,i+1} + \left( \sclcal{v}(\scl{X}^*_{s+1} ,\theta) - \sclcal{v}(\scl{Y}^*_{s+1} ,\theta) \right) \alpha_i  &\mbox{for} \ 1 \leq i \leq q-1 \\  
      \left( \sclcal{v}(\scl{X}^*_{s+1} ,\theta) - \sclcal{v}(\scl{Y}^*_{s+1} ,\theta) \right)\alpha_i  &\mbox{for} \ i = q\,, 
    \end{cases}
  \end{split}
\end{equation}
with
\begin{equation*}
  \begin{split}
    \scl{X}^*_{s+1} &= (z-\nu_2)\lambda + \scl{B}^*_{s+1,1}-\nu_1 + \frac{\frac{1}{2}(z-\nu_2)^2 + \gamma^2\scl{C}^*_{s+1,1} - 2\scl{C}^*_{s+1,1} \gamma (z-\nu_2)}{1-2\scl{C}^*_{s+1,1}}\,,\\
    \scl{Y}^*_{s+1} &= -\nu_2\lambda -\nu_1 + \frac{1}{2}\nu_2^2 \,,
  \end{split}
\end{equation*}
and terminal conditions $\scl{A}^*_T = \scl{B}^*_{T,i} = \scl{C}^*_{T,j} = 0$ for $i=1,\ldots,p$ and $j=1,\ldots,q$.
\label{cor:MGFQ}
\end{cor}
Proof: See Appendix \ref{app:LHARG}.\\

The derivation of the no-arbitrage condition for LHARG readily follows from Proposition~\ref{th:NOARBITRAGE}.
\begin{cor}
\label{cor:LHARGNOARBITRAGE}
  The LHARG model defined by eq.s~(\ref{eq:lrdyn}) and~(\ref{thetalharg}) with SDF specified as in~(\ref{LHARGSDF}) satisfies the no-arbitrage condition if, and only if
  \begin{equation}
    \nu_2 = \lambda + \frac{1}{2}. 
  \end{equation}
\end{cor}
Proof: See Appendix \ref{app:LHARG}.\\

To derive the price of vanilla options, for example, it is sufficient to know the MGF under the risk-neutral measure $\mathbb{Q}$ which has been given in Corollary~\ref{cor:MGFQ}. However, for exotic instruments it is essential to know the log-return dynamics under $\mathbb{Q}$. The comparison of the physical and risk-neutral MGFs provides us the one-to-one mapping among the parameters which trasforms the dynamics under $\mathbb{Q}$ into the dynamics under $\mathbb{P}$. 
\begin{prop}
\label{prop:RNDYNAMICS}
  Under the risk-neutral measure $\mathbb{Q}$ the realized variance still follows a LHARG process with parameters
  \begin{equation}
    \label{parQ}
    \begin{array}{lll}
      \beta^*_{d}  = \frac{1}{1-\theta \scl{Y}^*}\beta_{d}\,,  & \beta^*_{w}  = \frac{1}{1-\theta \scl{Y}^*}\beta_{w}\,,  & \beta^*_{m}  = \frac{1}{1-\theta \scl{Y}^*}\beta_{m}\,,\\
      \alpha^*_{d} = \frac{1}{1-\theta \scl{Y}^*}\alpha_{d}\,, & \alpha^*_{w} = \frac{1}{1-\theta \scl{Y}^*}\alpha_{w}\,, & \alpha^*_{m} = \frac{1}{1-\theta \scl{Y}^*}\alpha_{m}\,,\\
      \theta^*     = \frac{1}{1-\theta \scl{Y}^*}\theta\,,     & \delta^* = \delta\,, & \gamma^* = \gamma + \lambda + \frac{1}{2}\,,\\
      d^*          = \frac{1}{1-\theta \scl{Y}^*}d\,, &  &
    \end{array}
  \end{equation}
  where $\scl{Y}^* = - \lambda^2/2 - \nu_1 + \frac{1}{8}$.
\end{prop}
Proof: See Appendix \ref{app:RNDYNAMICS}.\\

From the previous results we can write the simplified risk-neutral MGF which allows us to reduce the computational burden when computing the backward recurrences.   
\begin{cor}
  Under $\mathbb{Q}$, the MGF for the LHARG model has the same form as in~(\ref{Pmgf})-(\ref{coefmgfP}) with equity risk premium $\lambda^* = -0.5$ and $d^*$, $\delta^*$, $\theta^*$, $\gamma^*$, $\alpha^*_l$, $\beta_l^*$ for $l=d,w,m$ as in~(\ref{parQ}).
\end{cor}

\subsection{Particular cases}

We now discuss two special cases of the model presented in the previous section. 
The first instance is the HARG model with Parabolic Leverage (P-LHARG) that we obtain setting $d=0$ in~(\ref{thetalharg}), while the second model is a LHARG with zero-mean leverage (ZM-LHARG). The shape of the leverage in the latter has been inspired by the model of~\cite{ChrJacOrnWan08} but in the present context it is enriched by a heterogeneous structure
\begin{equation*}
  \begin{split}
    \bar{\scl{\ell}}^{(d)}_t &= \epsilon_t^2 - 1 - 2\epsilon_t \gamma \sqrt{\mathrm{RV}_{t}}\,,\\
    \bar{\scl{\ell}}^{(w)}_t &= \frac{1}{4}\sum_{i=1}^4 \left( \epsilon_{t-i}^2 - 1 - 2\epsilon_{t-i} \gamma \sqrt{\mathrm{RV}_{t-i}} \right)\,,\\
    \bar{\scl{\ell}}^{(m)}_t &= \frac{1}{17}\sum_{i=5}^{21} \left(  \epsilon_{t-i}^2 - 1 - 2\epsilon_{t-i} \gamma \sqrt{\mathrm{RV}_{t-i}} \right)\,.
  \end{split}
\end{equation*}
The linear $\Theta(\mtx{RV}_t,\mtx{L}_t)$ in this case reads 
\begin{equation}
  \label{eq:ZM-LHARGtheta}
  \beta_d \mathrm{RV}_t^{(d)} + \beta_w \mathrm{RV}_t^{(w)} + \beta_m \mathrm{RV}_t^{(m)} + \alpha_d \bar{\scl{\ell}}_t^{(d)} + \alpha_w \bar{\scl{\ell}}_t^{(w)} + \alpha_m \bar{\scl{\ell}}_t^{(m)}\,, 
\end{equation}
which can be reduced to the form~(\ref{thetalharg}) setting $d = -(\alpha_d + \alpha_w + \alpha_m)$, $\beta_{l} = \beta_{l} - \alpha_{l} \gamma^2$ for $l=d,w,m$. 
As will be more clear in the following section, the introduction of the less constrained leverage allows the process to explain a larger fraction of the skewness and kurtosis observed in real data. However, similarly to what has been discussed in Section~\ref{sec:MFV} about~\cite{ChrJacOrnWan08}, it is no more guaranteed that the non centrality parameter of the gamma distribution is positive definite. Nonetheless, in the next section we will provide numerical evidence of the effectiveness of our analytical results in describing a regularized version of this model.

\subsection{Estimation and statistical properties}
\label{sec:ESTIMATION-STAT}

The estimation of the parameters characterizing the LHARG-RV family is greatly simplified by the use of Realized Volatility, which avoids any filtering procedure related to latent volatility processes. We compute the RV from tick-by-tick data for the S\&P 500 Futures, from January 1, 1990 to December 31, 2007. As pointed out in~\cite{CorFusLav12}, the choice of an adequate RV estimator is mandatory to reconcile the properties of LHARG-RV models with the realized volatility dynamics. Although we aknowledge the importance of the jump contribution in log-returns and realized volatility\footnote{See for instance~\cite{AndBolDie07}, \cite{CorPirRen10}, \cite{BS2006}}, these components are not included in the class of models considered in this paper. In order to exclude the effect of jump on log-return and volatility process, from the empirical analysis, we employ the same methodology adopted by \cite{CorFusLav12}: i) we estimate the total variation of the log-prices using the Two-Scale estimator proposed by~\cite{Zhang05}; ii) purify it from the jump component in prices by means of the Threshold Bipower variation method introduced in~\cite{CorPirRen10}; iii) remove the most extreme observations (jumps) in the volatility series. Finally, to overcome the problem of neglecting the contribution to the volatility due to the overnight effect we rescale our RV estimator to match the unconditional mean of the squared close-to-close daily returns. Further details about the construction of the RV measure are given in~\cite{CorFusLav12}.\\

The use of an RV proxy for the unobservable volatility allows us simply to employ a Maximum Likelihood Estimator (MLE) on historical data. Arguing as in~\cite{GouJas06}, the conditional transition density for the LHARG-RV family is available in closed-form, and so the log-likelihood reads  
\begin{equation*}
  \begin{split}
    &l_t^T (\delta,\theta,d,\beta_{d},\beta_{w},\beta_{m},\alpha_{d},\alpha_{w},\alpha_{m},\gamma)=\\ 
    &- \sum_{t=1}^T \left( \frac{\mathrm{RV}_t}{\theta} + \Theta\left(\mtx{RV}_{t-1},\mtx{L}_{t-1}\right)\right)
    + \sum_{t=1}^{T} \log \left( \sum_{k=1}^{\infty} \frac{\mathrm{RV}_t^{\delta+k-1}}{\theta^{\delta + k}\Gamma(\delta+k)}\frac{\Theta\left(\mtx{RV}_{t-1},\mtx{L}_{t-1}\right)^k}{k!} \right) 
  \end{split}
\end{equation*} 

where $\Theta\left(\mtx{RV}_{t-1},\mtx{L}_{t-1}\right)$ is given in eq.~(\ref{thetalharg}). To implement the MLE, we truncate the infinite sum on the right hand side to the 90$th$ order as done in~\cite{CorFusLav12}. Finally, the estimation of the market price of risk $\lambda$ in the log-return eq.~(\ref{RVdyn}) is performed regressing the centred and normalized log-returns on the realized volatility, in a similar way to eq.~(18) in~\cite{CorFusLav12}. As a proxy for the risk-free rate $r$ we employ the FED Fund rate.
\begin{center}
  \begin{table}[ht!]
    \begin{center}
      \begin{footnotesize}
	\begin{tabular}{lcccc}
	  &   \multicolumn{4}{c}{ Model }                   \\
	  \cline{2-5}   
	  &           &         &             &                  \\
	  Parameter         &    HARG                     &    HARGL                       &  P-LHARG                         &  ZM-LHARG                        \\
	  \cline{1-5}   
	  &           &         &             &                  \\
	  $\lambda$           &   \multicolumn{4}{c}{ 2.005}                   \\
	  &   \multicolumn{4}{c}{ (1.489) }                   \\
	  \cline{2-5}                    
	  $\theta$                 & 1.149e-005      & 1.116e-005      &1.068e-005         &1.117e-005         \\ 
	  & (1.036e-007)& (9.864e-008)  &(9.466e-008)   &(9.484e-008)   \\ 
	  $\delta$     	       & 1.358   & 1.395      &1.243         &1.78         \\ 
	  & (0.04566)& (0.04646)  &(0.0482)   &(0.04319)   \\ 
	  $\beta_{d} $	       & 3.959e+004    & 2.993e+004      &2.429e+004         &3.382e+004         \\ 
	  & (619.9)& (1037)  &(439.4)   &(180.1)   \\ 
	  $\beta_{w}$ 	       & 2.451e+004    & 2.796e+004      &2.317e+004         &2.542e+004         \\ 
	  & (1770)& (1247)  &(1199)   &(225)   \\ 
	  $\beta_{m}$          & 1.012e+004    &1.132e+004       &1.322e+004         &1.338e+004         \\ 
	  & (1644)&(897)   &(1690)   &(142.7)   \\ 
	  $\alpha_{d} $ 	   &        -                    &1.389e+004       &0.2376         &0.3991         \\ 
	  &                             &(1235)   &(0.00113)   &(0.007164)   \\ 
	  $\alpha_{w} $	       &        -                    &        -                       &0.1194         &0.3446         \\ 
	  &                             &                                &(0.002058)   &(0.01162)   \\ 
	  $\alpha_{m} $	       &        -                    &        -                       &3.85e-006         &0.4034         \\ 
	  &                             &                                &(3.649e-006)   &(0.02082)   \\ 
	  $\gamma $            &        -                    &        -                       &223.7         &134.8         \\ 
	  &                             &                                &(5.122)   &(9.525)   \\ 
	  &                             &                                &                                &                                   \\ 
	  &                             &                                &                                &                                   \\ 
	  $\nu_1$            & -2794   & -3119     & -3069      & -3375      \\ 
	  Log-likelihood   	   & -25344     &  -25279      &  -25234      & -25172       \\ 
	  Persistence      	   &0.8532   & 0.8495    &0.8391     &0.8116    \\
	  &           &         &             &                  \\
	  \hline
	\end{tabular}
      \end{footnotesize}
    \end{center}
    \hspace{-1.5cm}
    \vspace{1cm}
    \caption{Maximum likelihood estimates, robust standard errors, and models' performance.
      The historical data for the HARG, HARGL, P-LHARG and ZM-LHARG models are given by the daily RV measure computed on 
      tick-by-tick data for the S\&P500 Futures (see Section~\ref{sec:ESTIMATION-STAT}). 
      For all three models, the estimation period ranges from 1990-2005. The parameter $\nu_1$ for each model 
    has been fitted on option prices.}
    \label{tab:estimates}
  \end{table}
\end{center}

In Table~\ref{tab:estimates} we report the parameter values estimated via maximum likelihood for four different models, HARG, HARGL, P-LHARG, and ZM-LHARG\footnote{In~\cite{CorFusLav12} log-returns were expressed on a daily and percentage basis, whilst the realized volatility was on a yearly and percentage basis. Here, both log-returns and volatilities are on a daily and decimal basis.}. We also show the parameter standard deviations (in parenthesis), and the value of the log-likelihood. All parameters are statistically significant except the monthly leverage component of P-LHARG. As already documented in~\cite{Cor09} and~\cite{CorFusLav12} the RV coefficients show a decreasing impact of the past lags on the present value of the RV. As far as the leverage components are concerned there is no evidence of a clear relation among different lags. Finally, it is worth noting that the inclusion of leverage with heterogeneous structure improves the likelihood of competitor HARG and HARGL models. 

While we can ensure that P-LHARG model satisfies condition (\ref{eq:poscon}), for the ZM-LHARG model  the relation ~(\ref{eq:ZM-LHARGtheta}) cannot be prevented from obtaining negative values. Since the ZM-LHARG is worth considering, we provide some numerical evidence supporting the analytical MGF as a reliable approximation of the MGF computed by simulation. We compare an extensive Monte Carlo (MC) simulation of the ZM-LHARG dynamics where the non centrality parameter is artificially bounded from below (by zero) with the analytical MGF computed according to Proposition~\ref{prop:mgfP}. 
As the probability of obtaining a negative value for the non centrality of the gamma distribution is small (given the parameter values in Table~\ref{tab:estimates}),
we can assess that the analytical MGF is a good approximation of the unknown MGF of the regularized ZM-LHARG. We fix the number of MC to $0.5\times 10^6$ and consider six relevant maturities, one day $(T=1)$, one week $(T=5)$, one month $(T=22)$, one quarter $(T=63)$, six months $(T=126)$, and one year $(T=256)$. 
\begin{figure}
  \begin{center}
    \includegraphics[width=0.85\textwidth]{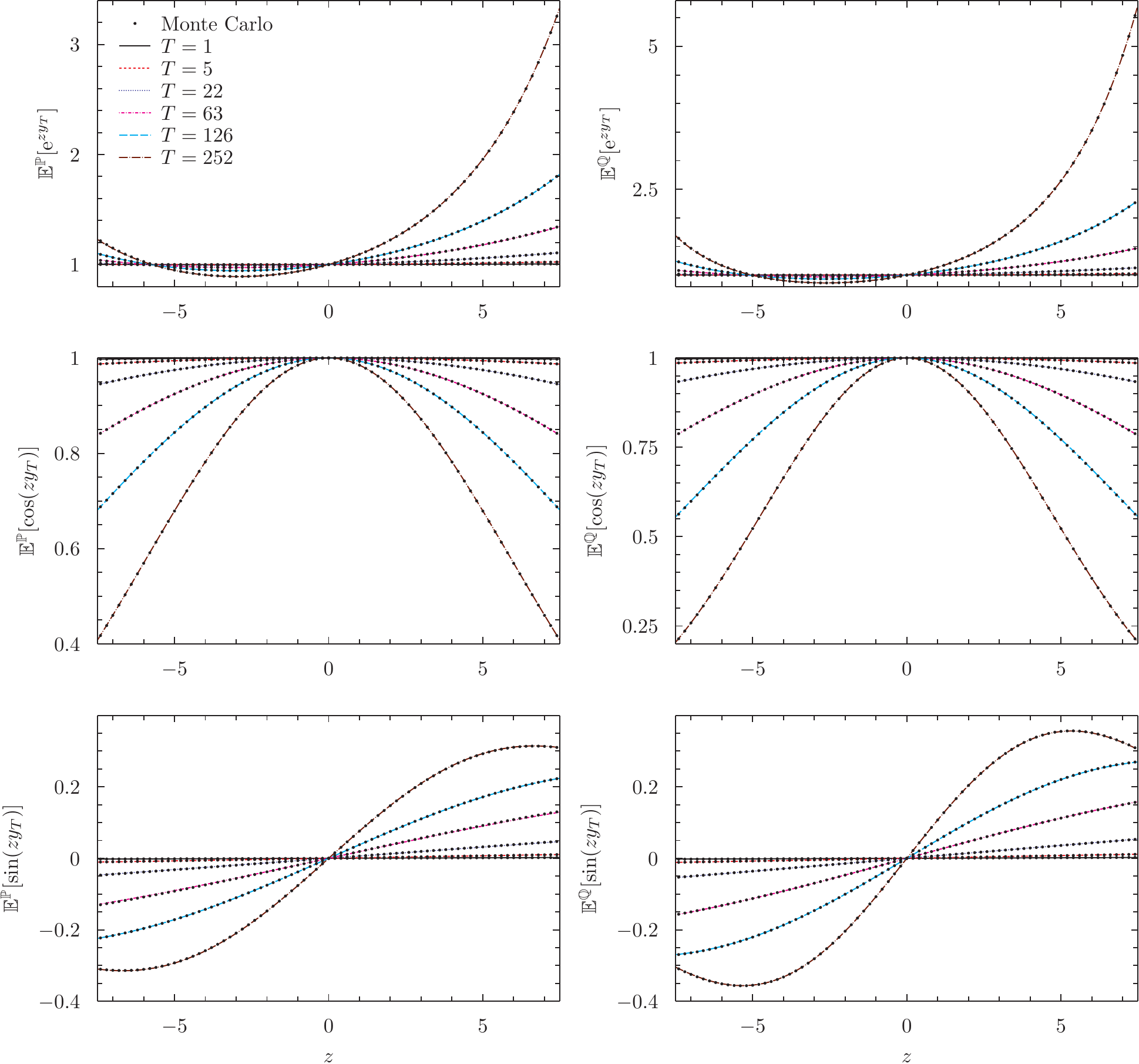}
    \caption{\label{fig:LHARG_CJOW} 
    Left column, from top to bottom: MGF, real and imaginary parts of the characteristic function of the ZM-LHARG process under the physical measure $\mathbb{P}$. Right column, from top to bottom: MGF, real and imaginary parts of the Characteristic Function of the ZM-LHARG process under the risk-neutral measure $\mathbb{Q}$. The lines correspond to different maturities $T=1,5,22,63,126,252$, while points to Monte Carlo expected values; Monte Carlo error bars are smaller than the point size. 
  }
\end{center}
\end{figure}
In the left column from top to bottom of Figure~\ref{fig:LHARG_CJOW} we plot the MGF, the real and imaginary parts of the characteristic function under the physical measure, respectively, while in the right column we show the same quantities under the risk-neutral measure. The lines correspond to the analytical MGFs while the MC expectations are represented by points whose size is larger than the associated error bars. The quality of the agreement is extremely high. Moreover, the MC estimate of the probability associated with the event $\Theta(\mtx{RV}_{t-1},\mtx{L}_{t-1})<0$ is $2\times 10^{-5}$ under $\mathbb{P}$, and $3\times 10^{-6}$ under $\mathbb{Q}$, confirming once more the reliability of the approximation.
\begin{figure}
  \begin{center}
    \includegraphics[width=0.85\textwidth]{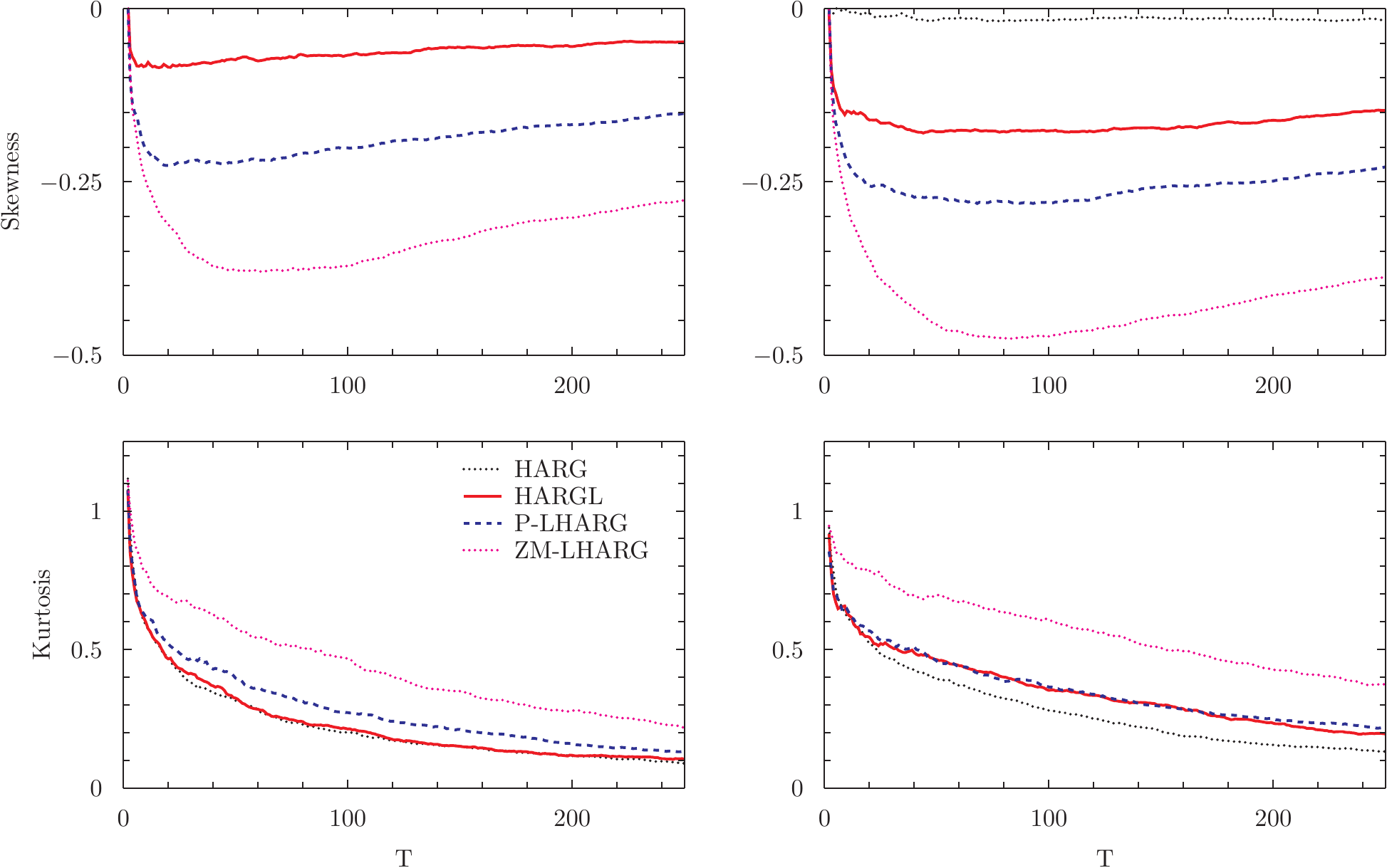}
    \caption{\label{fig:CUMULANTS} 
    Left column, from top to bottom: skewness and excess kurtosis of the HARG, HARGL, P-LHARG, and ZM-LHARG processes under the physical measure $\mathbb{P}$. Right column: same as the left column, but under the risk-neutral measure. 
  }
\end{center}
\end{figure}

Crucial ingredients for reproducing the shape of the implied volatility surface are the term structure of skewness and kurtosis generated by a given 
option pricing model.
Therefore, in Figure~\ref{fig:CUMULANTS} we compare the skewness and excess kurtosis associated to the four models HARG, HARGL, P-LHARG, and ZM-LHARG. 
We do not show the skewness for the HARG case under $\mathbb{P}$, since this model is not designed to explain the negative skewness. 
When moving to $\mathbb{Q}$, the genuine effect of the calibration of $\nu_1$ is to induce a small negative skewness. 
It is worth noticing that for the LHARG-RV models adding the heterogeneous components not only improves the skewness upon the HARGL model, 
but also considerably increases the excess kurtosis. 
As far as under the $\mathbb{Q}$ measure is concerned, the HARGL process catches up to the P-LHARG model both in terms of skewness and kurtosis, 
while the ZM-LHARG always outperforms all the competitor models. 

\section{Valuation performance}
\label{sec:Valuation}

\subsection{Option pricing methodology}

We apply the same option pricing procedure for both LHARG models, based on change of measure described by (\ref{LHARGSDF}), and MGF formula given by (\ref{Pmgf})-(\ref{coefmgfP}). To derive risk-neutral dynamics we need to fix parameters of SDF, $\nu_1$ and $\nu_2$. While the latter is determined by the no-arbitrage condition (Proposition \ref{cor:LHARGNOARBITRAGE}), the former has to be calibrated on option prices. Following the same reasoning as~\cite{CorFusLav12}, we perform the unconditional calibration of $\nu_1$ such that the model generated and the average market IV for a one-year time to maturity at-the-money option coincide.\\

We employ the option pricing numerical method termed COS, introduced by~\cite{FO08}, and which has been proven to be efficient. The method is based on Fourier-cosine expansions and is available as long as the characteristic function of log-returns is known. The numerical algorithm exploits the close relation of the characteristic function with the series coefficients of Fourier-cosine expansion of the density function.\\

To sum up, we proceed pricing options following four steps: (i) estimation under the physical measure $\mathbb{P}$, (ii) unconditional calibration of the parameter $\nu_1$ (iii) mapping of the parameters of the model estimated under $\mathbb{P}$ into the parameters under $\mathbb{Q}$, and (iv) approximation of option prices by the COS method using the MGF formula in~(\ref{Pmgf})-(\ref{coefmgfP}) with parameters under measure $\mathbb{Q}$.\\

\subsection{Results}

In this section we present empirical results for option pricing with LHARG models. For the sake of completeness we also compare LHARG models with the HARG model with no leverage and with the HARGL presented in~\cite{CorFusLav12}. Since the functional form of the leverage of the latter model is not consistent with the current general framework, closed-form formulae for the MGF and for option pricing are not available. Thus, we resort to numerical methodologies such as extensive Monte Carlo scenario generation.\\

We perform our analysis on European options, written on the S$\&$P 500 index. The time series of option prices range from January 1, 1996 to December 31, 2004 and the data are downloaded from OptionMetrics. As is customary in the literature (see~\cite{GBAEM08}), we filter out options with time to maturity less than 10 days or more than 365 days, implied volatility larger than $70 \%$, and prices less than 5 cents. Following ~\cite{CorFusLav12}, we consider only out-of-the-money (OTM) put and call options for each Wednesday. Moreover we discard deep out-of-the-money options (moneyness larger than $1.2$ for call options and less than $0.8$ for put options). The procedure yields a total of $41536$ observations.\\

As a measure of the option pricing performance we use the percentage Implied Volatility Root Mean Square Error ($RMSE_{IV}$) put forward by~\cite{Ren97} and computed as
\begin{equation*}
  RMSE_{IV} = \sqrt{\frac{1}{N}\sum_{i=1}^N \left( IV^{mkt}_i - IV^{mod}_i \right)^2} \times 100\,,
\end{equation*}
where $N$ is the number of options, $IV^{mkt}$ and  $IV^{mod}$ represent the market and model implied volatilities, respectively.  An alternative performance measure corresponds to the Price Root Mean Square Error ($RMSE_{P}$) defined in a similar way as $RMSE_{IV}$ but with implied volatilities replaced by relative prices. We employ the $RMSE_{IV}$ measure since it tends to put more weight on OTM options, while the $RMSE_{P}$ emphasizes the importance of ATM options.\\
%
%%%%%%%%%%%%%%%%%%%%%%%%%%%%%%%%%%%%%%%%%%%%%%%%%%%%%%%%%%%%%%%%%%%%%%%%%%%%%%%%%%%%%%%%%%%%%%%%%%%%%%%%%
% Tab Overall Performance
%%%%%%%%%%%%%%%%%%%%%%%%%%%%%%%%%%%%%%%%%%%%%%%%%%%%%%%%%%%%%%%%%%%%%%%%%%%%%%%%%%%%%%%%%%%%%%%%%%%%%%%%%
\begin{center}
  \begin{table}[ht!]
    \begin{center}
      \begin{footnotesize}
	\begin{tabular}{lcccc}
	  \multicolumn{5}{c}{Implied Volatility RMSE} \\ 
	  &  &  &  &    \\
	  &  & \multicolumn{3}{c}{Moneyness} \\ 
	  \cline{3-5}
	  Model  & & \multicolumn{1}{c}{$ 0.9 < m < 1.1 $ } &  & \multicolumn{1}{c}{$ 0.8 < m < 1.2 $} \\ 
	  \hline 
	  HARGL    &  &3.817  &  & 6.103 \\
	  \rowcolor{gray!30} P-LHARG/HARGL   &  &0.960 &  & 0.824 \\
	  ZM-LHARG/HARGL   &  &0.927 &  & 0.775 \\
	  \rowcolor{gray!30} P-LHARG/HARG   &  &0.891 &  & 0.746 \\
	  ZM-LHARG/HARG   &  &0.861 &  & 0.702 \\
	  \rowcolor{gray!30} ZM-LHARG/P-LHARG   &  &0.966 &  & 0.942 \\
	  \hline
	\end{tabular}
      \end{footnotesize}
    \end{center}

    \vspace{1cm}
    \caption{Global option pricing performance on S\&P500 out-of-the-money options from January 1, 1996 to December 31, 2004, computed with the RV measure estimated from 1990 to 2007.\newline We use the maximum likelihood parameter estimates from Table \ref{tab:estimates}. First row: percentage implied volatility root mean squared error ($RMSE_{IV}$) of the HARGL model (benchmark) for different moneyness range.Second and subsequent rows: relative $RMSE_{IV}$ of the selected models.}
    \label{tab:performance_main}
  \end{table}
\end{center} 
The result of our empirical analysis is that both LHARG models outperform competing RV-based stochastic volatility models (HARG, HARGL). Table \ref{tab:performance_main} shows that P-LHARG outperforms HARG and HARGL by about $11 \%$ and $4 \%$, respectively, in range of moneyness $0.9 < m < 1.1$ and by about $35 \%$ and $17 \%$, respectively, in range of moneyness $0.8 < m  < 1.2$. ZM-LHARG outperforms HARG and HARGL by about $14 \%$ and $7 \%$, respectively, in range of moneyness $0.9<m<1.1$ and by about $30 \%$ and $22 \%$, respectively, in range of moneyness $0.8<m<1.2$. ZM-LHARG improves P-LHARG by about $3 \%$ and $6 \%$, in range of moneyness $0.9<m<1.1$ and $0.8<m<1.2$, respectively.\\

The detailed analysis in Table \ref{tab:performance_detail_IV} confirms that the main advantage of LHARG models is the ability to capture the volatility smile. While the performance of all models in the at-the-money region is similar, both LHARG models significantly outperform HARG and HARGL in the range of moneyness  $1.1 < m < 1.2$ and even more at the put side region $0.8 < m < 0.9$. This improvement stems from the higher flexibility of the model obtained with a multi-component leverage structure.\\
%
%%%%%%%%%%%%%%%%%%%%%%%%%%%%%%%%%%%%%%%%%%%%%%%%%%%%%%%%%%%%%%%%%%%%%%%%%%%%%%%%%%%%%%%%%%%%%%%%%%%%%%%%%%%%%%%%%
% Tab Performance detail IV
%%%%%%%%%%%%%%%%%%%%%%%%%%%%%%%%%%%%%%%%%%%%%%%%%%%%%%%%%%%%%%%%%%%%%%%%%%%%%%%%%%%%%%%%%%%%%%%%%%%%%%%%%%%%%%%%%
\begin{center}
  \begin{table}[tbp!]
    \begin{center}
      \begin{scriptsize} 
	\begin{tabular}{lccccccc}
	  & \multicolumn{7}{c}{Maturity} \\ \cline{2-8}
	  &  &  &  &  &  &  &  \\
	  Moneyness & \multicolumn{1}{c}{$\tau \leq 50 $ } &  & \multicolumn{1}{c}{$ 50 < \tau \leq 90$} &  & \multicolumn{1}{c}{ $ 90  < \tau \leq 160 $} &  & \multicolumn{1}{c}{$ 160 < \tau$} \\ 
	  \hline 
	  & && && && \\ 
	  Panel A & \multicolumn{7}{c}{HARGL Implied Volatility RMSE} \\ 
	  & && && && \\
	  \rowcolor{gray!30}$0.8 \leq m \leq0.9$ &16.140&& 8.267&& 6.803&& 5.516 \\
	  $0.9 < m \leq 0.98$ &5.598&& 4.411&& 3.939&& 3.872 \\
	  \rowcolor{gray!30}$0.98 < m \leq 1.02$ &2.681&& 2.780&& 2.872&& 3.261\\ 
	  $1.02 < m \leq 1.1$ &3.061&& 2.783&& 2.789&& 3.070\\ 
	  \rowcolor{gray!30}$1.1< m \leq 1.2$ &5.412&& 3.262&& 3.217&& 3.206\\ 
	  & && && && \\
	  Panel B &\multicolumn{7}{c}{P-LHARG/HARGL Implied Volatility RMSE} \\
	  & && && && \\
	  \rowcolor{gray!30}$0.8 \leq m \leq 0.9$ &0.691&& 0.917&& 0.939&& 0.973\\ 
	  $0.9 < m \leq 0.98$ &0.892&& 0.925&& 0.975&& 1.020\\ 
	  \rowcolor{gray!30}$0.98 < m \leq 1.02$ &0.988&& 1.025&& 1.071&& 1.087\\ 
	  $1.02 < m \leq 1.1$ &0.975&& 1.069&& 1.120&& 1.114\\ 
	  \rowcolor{gray!30}$1.1< m \leq 1.2 $ &0.799&& 0.949&& 0.975&& 1.048\\ 
	  & && && && \\
	  Panel C &\multicolumn{7}{c}{ZM-LHARG/HARGL Implied Volatility RMSE} \\
	  & && && && \\
	  \rowcolor{gray!30}$0.8 \leq m \leq 0.9$ &0.648&& 0.824&& 0.844&& 0.902\\ 
	  $0.9 < m \leq 0.98$ &0.841&& 0.870&& 0.928&& 1.001\\ 
	  \rowcolor{gray!30}$0.98 < m \leq 1.02$ &0.988&& 1.035&& 1.073&& 1.096\\ 
	  $1.02 < m \leq 1.1$ &0.961&& 1.041&& 1.089&& 1.101\\ 
	  \rowcolor{gray!30}$1.1< m \leq 1.2 $ &0.784&& 0.849&& 0.854&& 0.972\\ 
	  & && && && \\
	  Panel D &\multicolumn{7}{c}{P-LHARG/HARG Implied Volatility RMSE} \\
	  & && && && \\
	  \rowcolor{gray!30}$0.8 \leq m \leq 0.9$ &0.616&& 0.825&& 0.847&& 0.890\\ 
	  $0.9 < m \leq 0.98$ &0.802&& 0.852&& 0.909&& 0.972\\ 
	  \rowcolor{gray!30}$0.98 < m \leq 1.02$ &0.965&& 1.003&& 1.045&& 1.062\\ 
	  $1.02 < m \leq 1.1$ &0.934&& 1.007&& 1.060&& 1.065\\ 
	  \rowcolor{gray!30}$1.1< m \leq 1.2 $ &0.836&& 0.831&& 0.857&& 0.942\\ 
	  & && && && \\
	  Panel E &\multicolumn{7}{c}{ZM-LHARG/HARG Implied Volatility RMSE} \\
	  & && && && \\
	  \rowcolor{gray!30}$0.8 \leq m \leq 0.9$ &0.577&& 0.741&& 0.761&& 0.825\\ 
	  $0.9 < m \leq 0.98$ &0.757&& 0.801&& 0.865&& 0.954\\ 
	  \rowcolor{gray!30}$0.98 < m \leq 1.02$ &0.965&& 1.013&& 1.047&& 1.070\\ 
	  $1.02 < m \leq 1.1$ &0.920&& 0.981&& 1.030&& 1.052\\ 
	  \rowcolor{gray!30}$1.1< m \leq 1.2$ &0.821&& 0.743&& 0.751&& 0.874\\ 
	  & && && && \\
	  Panel F &\multicolumn{7}{c}{ZM-LHARG/P-LHARG Implied Volatility RMSE} \\
	  & && && && \\
	  \rowcolor{gray!30}$0.8 \leq m \leq 0.9$ &0.937&& 0.898&& 0.899&& 0.927\\ 
	  $0.9 < m \leq 0.98$ &0.943&& 0.940&& 0.952&& 0.981\\ 
	  \rowcolor{gray!30}$0.98 < m \leq 1.02$ &1.000&& 1.010&& 1.002&& 1.008\\ 
	  $1.02 < m \leq 1.1$ &0.986&& 0.974&& 0.972&& 0.989\\ 
	  \rowcolor{gray!30}$1.1< m \leq 1.2 $ &0.982&& 0.894&& 0.876&& 0.927\\ 
	  & && && && \\
	  \hline
	\end{tabular}
      \end{scriptsize}
    \end{center}

    \vspace{1cm}
    \caption{Option pricing performance on S\&P500 out-of-the-money options from January 1, 1996 to December 31, 2004, computed with the RV measure estimated from 1990 to 2007.\newline We use the maximum likelihood parameter estimates from Table \ref{tab:estimates}. Panel A: percentage implied volatility root mean squared error ($RMSE_{IV}$) of the HARGL model sorted by moneyness and maturity.Panels B to F: relative $RMSE_{IV}$ sorted by moneyness and maturity.}
    \label{tab:performance_detail_IV}
  \end{table}
\end{center}
%%%%%%%%%%%%%%%%%%%%%%%%%%%%%%%%%%%%%%%%%%%%%%%%%%%%%%%%%%%%%%%%%%%%%%%%%%%%%%%%%%%%%%%%%%%%%%%%%%%%%%%%
%%% End Latex Table      %%%%%%%%%%%%%%%%%%%%%%%%%%%%%%%%%%%%%%%%%%%%%%%%%%%%%%%%%%%%%%%%%%%%%%%%%%%%%%%
%%%%%%%%%%%%%%%%%%%%%%%%%%%%%%%%%%%%%%%%%%%%%%%%%%%%%%%%%%%%%%%%%%%%%%%%%%%%%%%%%%%%%%%%%%%%%%%%%%%%%%%%

Panel B of Table \ref{tab:performance_detail_IV} compares the performance of HARGL and P-LHARG. It shows the advantage of heterogeneous leverage compared to one-day binary leverage. Improvement for short maturities and moneyness $0.8<m<0.9$ reaches  about $30 \%$. For longer maturities and moneyness below $0.9$,  P-LHARG still outperforms HARGL, obtaining $3\%-8 \%$ smaller $RMSE_{IV}$. In the other moneyness regions the two models perform quite similarly.\\
  
The ratio between $RMSE_{IV}$  of HARGL and ZM-LHARG is displayed in Panel C of Table \ref{tab:performance_detail_IV}. The advantage of zero-mean heterogeneous leverage over one-day binary leverage is even stronger than in the case of P-LHARG. For all deep out-of-the-money options, the error generated by ZM-LHARG on implied volatility is smaller than in the case of HARGL. For short maturities and moneyness less than $0.9$ we obtain about $35 \%$ improvement. ZM-LHARG also performs better for deep out-of-the-money options on call side ($1.1<m<1.2$), where improvement varies from $3 \%$ to $22 \%$.\\  

Comparing model HARG without leverage with P-LHARG and ZM-LHARG in Panel D and Panel E, respectively, the superiority of the latter two is even more apparent. While the performance for the ATM options is comparable, the OTM options models with heterogeneous leverage generate considerable improvement over models without leverage. In the extreme case of OTM short maturity put options P-LHARG and ZM-LHARG produces errors which are smaller by $38\%$ - $42\%$, respectively.\\

The last Panel (F) of Table \ref{tab:performance_detail_IV} compares ZM-LHARG with P-LHARG. It shows that the ability of ZM-LHARG model to reproduce higher levels of skewness and kurtosis, permits this more flexible model to outperform the more constrained P-LHARG model. The outperformance is systematic, from ATM options, where $RMSE_{IV}$ is essentially the same, to deep out-of-the-money ($m>1.1$ or $m<0.9$) where $RMSE_{IV}$ is smaller by about $10\%$.\\

To summarize, the proposed LHARG models are better able to reproduce the IV level for OTM options, improving upon the considered HARG and HARGL models. The heterogeneous structure of the leverage thus appears to be a necessary ingredient for more accurate modeling of the IV smile.

\section{Conclusions}
\label{sec:Conclusions}

In this paper we propose a very general framework which includes a wide class of discrete time models featuring multiple components structure in both volatility and leverage and a flexible pricing kernel with multiple risk premia. Within this framework we characterize the recursive formulae for the analytical MGF under $\mathbb{P}$ and $\mathbb{Q}$, the change of measure obtained using a flexible exponentially affine SDF, and the analytical no-arbitrage conditions. Then, we focus on a specific new class of realized volatility models, named LHARG, which extend the HARGL model of~\citet{CorFusLav12} to incorporate analytically tractable heterogeneous leverage structures with multiple components. This feature allows for higher skewness and kurtosis which enables LHARG models to outperform other RV-based stochastic volatility models (HARG, HARGL) in pricing out-of-the-money options. The proposed general framework can be employed to include several additional features like jumps in log-return and realized volatility, overnight effect, combination of GARCH and realized volatility models, and switching in volatility regimes.

\section*{Acknowledgements}
All authors warmly thank Nicola Fusari and Davide La Vecchia for helpful comments and fruitful discussions and anonymous referees for helpful remarks.  Additionally, we would like to thank Peter Lieberman for the help in text editing.

%\footnotesize
%\singlespacing
%\bibliographystyle{elsarticle-harv}
%\bibliographystyle{Chicago}
%\bibliography{biblioHARG}

%\normalsize
%\onehalfspacing

\appendix
\section{No arbitrage condition}
\label{app:NOARBITRAGE}
The no-arbitrage conditions are
\begin{equation}
  \EP{M_{s,s+1} | \mathcal{F}_s} = 1 \ \ \mbox{for} \ s \in \mathbb{Z}_+, 
\end{equation}

\begin{equation}
  \label{NAcon2}
  \EP{M_{s,s+1} \re^{y_{s+1}}| \mathcal{F}_s} = \re^r \ \ \mbox{for} \ s \in \mathbb{Z}_+.
\end{equation}

The first condition is satisfied by definition of $M_{s,s+1}$. Before moving to the second condition, let us rewrite the SDF as 
\begin{equation}
  \label{sdfform2}
  \begin{split}
    M_{s,s+1} &= \frac{\re^{-\vct{\nu_{1}}\cdot\vct{F}_{s+1}   - \nu_{2} y_{s+1} }}{\EP{\re^{-\vct{\nu_{1}}\cdot\vct{F}_{s+1} - \nu_{2} y_{s+1}} | \mathcal{F}_{s}}}\\
    &=  \exp \left( 
    \begin{split}
      &- \sclcal{A}(- \nu_{2},-\vct{\nu_{1}},\vct{0}) - \sum_{i=1}^p\vctcal{B}_{i}(- \nu_{2},-\vct{\nu_{1}},\vct{0})\cdot \vct{F}_{s+1-i} \\
      &- \sum_{i=1}^q\vctcal{C}_{i}(- \nu_{2},-\vct{\nu_{1}},\vct{0})\cdot\vct{\ell}_{s+1-i}-\vct{\nu_{1}}\cdot\vct{F}_{s+1} - \nu_{2} y_{s+1} 
    \end{split}
    \right),
  \end{split}
\end{equation}
where $\vct{\nu_1} = (\nu_1,\ldots,\nu_1)^t \in \mathbb{R}^k$ and functions $\sclcal{A}$, $\vctcal{B}_i$ and $\vctcal{C}_j$ are defined in~(\ref{affine}). Finally, the condition (\ref{NAcon2}) reads
\begin{equation}
  \begin{split}
    &\EP{\exp \left( -\boldsymbol{\vct{\nu}_{1}}\cdot\vct{F}_{s+1}  + \left( 1 - \nu_{2} \right) y_{s+1} \right)| \mathcal{F}_s}\\
    &=  \exp \left( r + \sclcal{A}(- \nu_{2},-\vct{\nu}_{1},\vct{0}) +\sum_{i=1}^p \vctcal{B}_{i}(-\nu_{2},-\vct{\nu}_{1},\vct{0})\cdot\vct{F}_{s+1-i} + \sum_{j=1}^q\vctcal{C}_{j}(-\nu_{2},-\vct{\nu}_{1},\vct{0})\cdot\vct{\ell}_{s+1-j}\right).
  \end{split}
\end{equation}
Using once again the relation~(\ref{affine}) we obtain the no-arbitrage conditions.

\section{Computation of MGF}
\label{app:MGFCOMPUTATION}
\label{th:GF-MGFQ}
Under the risk-neutral measure $\mathbb{Q}$ the MGF of the log-returns $y_{t,T} = \log(S_T/S_t)$ conditional on the information available at time $t$ is of the form
\begin{equation}
\label{eq:GF-MGFQ}
\MGFQ(t,T,z) = \re^{\scl{A}^*_{t}+\sum_{i=1}^{p}\vct{B}^*_{t,i}\cdot\vct{F}_{t+1-i}+\sum_{j=1}^{q}\vct{C}^*_{t,j}\cdot\vct{\ell}_{t+1-j}}\ , \end{equation} 
where
\begin{equation}
  \label{eq:GF-MGFQrecursive}
  \begin{split}
    \scl{A}^*_{s} &= \scl{A}^*_{s+1} + \sclcal{A}(z-\nu_2,\vct{B}^*_{s+1,1}-\vct{\nu}_{1},\vct{C}^*_{s+1,1})-\sclcal{A}(-\nu_2,-\vct{\nu}_{1},\vct{0})\\
    \vct{B}^*_{s,i} &= 
    \begin{cases}
      \vct{B}^*_{s+1,i+1} + \vctcal{B}_i(z-\nu_2,\vct{B}^*_{s+1,1}-\vct{\nu}_{1},\vct{C}^*_{s+1,1})-\vctcal{B}_i(-\nu_2,-\vct{\nu}_{1},\vct{0}) & \mbox{if} \ 1 \leq i \leq p-1\\
      \vctcal{B}_i(z-\nu_2,\vct{B}^*_{s+1,1}-\vct{\nu}_{1},\vct{C}^*_{s+1,1})-\vctcal{B}_i(-\nu_2,-\vct{\nu}_{1},\vct{0}) & \mbox{if} \  i = p
    \end{cases}\\
    \vct{C}^*_{s,j} &= 
    \begin{cases}
      \vct{C}^*_{s+1,j+1} + \vctcal{C}_j(z-\nu_2,\vct{B}^*_{s+1,1}-\vct{\nu}_{1},\vct{C}^*_{s+1,1})-\vctcal{C}_j(-\nu_2,-\vct{\nu}_{1},\vct{0}) & \mbox{if} \ 1 \leq j \leq q-1\\ 
      \vctcal{C}_j(z-\nu_2,\vct{B}^*_{s+1,1}-\vct{\nu}_{1},\vct{C}^*_{s+1,1})-\vctcal{C}_j(-\nu_2,-\vct{\nu}_{1},\vct{0}) & \mbox{if} \ j = q
    \end{cases}
  \end{split}
\end{equation}
and $\scl{A}_T^* = 0$, $\vct{B}^*_{T,i} = \vct{C}^*_{T,j} = \vct{0} \in \mathbb{R}^k$ for $i=1,\ldots,p$ and $j=1,\ldots,q$.\\

The above relation can be derived using the expression for the SDF given in~(\ref{sdfform2}), repeatedly and using the tower law of conditional expectation we obtain
\begin{equation*}
  \begin{split}
    &\MGFQ(t,T,z)\\ 
    &= \EQ{\re^{z y_{t,T} } | \mathcal{F}_{t}}\\
    &=\EP{M_{t,t+1}\ldots M_{T-1,T} \re^{z y_{t,T} } | \mathcal{F}_{t}}\\
    &=\EP{M_{t,t+1}\ldots M_{T-2,T-1}\re^{z y_{t,T-1} }\EP{M_{T-1,T}\re^{zy_T}| \mathcal{F}_{T-1}} | \mathcal{F}_{t}}\\
    &=\EP{
      \begin{split}
	&M_{t,t+1}\ldots M_{T-2,T-1}\re^{z y_{t,T-1}-\sclcal{A}(-\nu_{2},-\vct{\nu}_{1},\vct{0})-\sum_{i=1}^{p}\vctcal{B}_i(-\nu_{2},-\vct{\nu}_{1},\vct{0})\cdot\vct{F}_{T-i}}\\
	&\times\re^{- \sum_{j=1}^{q}\vctcal{C}_j(-\nu_{2},-\vct{\nu}_{1},\vct{0})\cdot\vct{\ell}_{T-i} }
	\EP{\re^{-\vct{\nu}_{1}\cdot\vct{F}_{T} + (z-\nu_{2}) y_{T}}|\mathcal{F}_{T-1}}
      \end{split} 
    | \mathcal{F}_{t}}\\
    &=\EP{
      \begin{split}
	&M_{t,t+1}\ldots M_{T-2,T-1}\re^{z y_{t,T-1} + \sclcal{A}(z-\nu_{2},-\vct{\nu}_{1},\vct{0})-\sclcal{A}(-\nu_{2},-\vct{\nu}_{1},\vct{0})}\\
	&\times\re^{\sum_{i=1}^{p}\left[\vctcal{B}_i(z- \nu_{2},-\vct{\nu}_{1},\vct{0})-\vctcal{B}_i(-\nu_{2},-\vct{\nu}_{1},\vct{0})\right]\cdot{\vct{F}}_{T-i}+\sum_{j=1}^{q}\left[\vctcal{C}_j(z-\nu_{2},-\vct{\nu}_{1},\vct{0})-\vctcal{C}_j(-\nu_{2},-\vct{\nu}_{1},\vct{0})\right]\cdot\vct{\ell}_{T-j}}
      \end{split} 
    | \mathcal{F}_{t}}\\
    &=\EP{
      M_{t,t+1}\ldots M_{T-2,T-1}\re^{z y_{t,T-1} + \scl{A}^*_{T-1}+\sum_{i=1}^{p}\vct{B}^*_{T-1,i}\cdot\vct{F}_{T-i}+\sum_{j=1}^{q}\vct{C}^*_{T-1,j}\cdot\vct{\ell}_{T-j}}
    | \mathcal{F}_{t}}\\
    &=\EP{
      \begin{split}
	&M_{t,t+1}\ldots M_{T-3,T-2}\re^{z y_{t,T-2} + \scl{A}^*_{T-1}}\\
	&\times\EP{M_{T-2,T-1}\re^{z y_{T-1}+\sum_{i=1}^{p}\vct{B}^*_{T-1,i}\cdot\vct{F}_{T-i}+\sum_{j=1}^{q}\vct{C}^*_{T-1,j}\cdot\vct{\ell}_{T-j} }|\mathcal{F}_{T-2}}
      \end{split}
    | \mathcal{F}_{t}}\\
    & = \ldots\\
    & = \re^{\scl{A}^*_{t}+\sum_{i=1}^{p}\vct{B}^*_{t,i}\cdot\vct{F}_{t+1-i}+\sum_{j=1}^{q}\vct{C}^*_{t,j}\cdot\vct{\ell}_{t+1-j}}\,.
  \end{split}
\end{equation*}

Finally, the MGF under $\mathbb{P}$ readily follows by noticing that for $\nu_1=\nu_2=0$ the SDF reduces to one, therefore $\MGFP(t,T,z)=\varphi^\mathbb{Q}_{00}(t,T,z)$. 
\section{MGF computation for LHARG and no-arbitrage conditions}
\label{app:LHARG}
Firstly, we derive the explicit form of the scalar functions $\sclcal{A}$, $\sclcal{B}_i$ and $\sclcal{C}_j$. In the case of LHARG we have $\cmp{F}{}_t=\mathrm{RV}_t$.
Then,
\begin{equation}
  \begin{split}
    &\EP{\re^{ z y_{s} + b \mathrm{RV}_{s} + c \scl{\ell}_{s} }| \mathcal{F}_{s-1}}\\
    &=\re^{zr}\EP{\re^{(z \lambda + b)\mathrm{RV}_{s}}\EP{\re^{z \sqrt{\mathrm{RV}_s} \epsilon_{s} + c (\epsilon_s - \gamma \sqrt{\mathrm{RV}_s})^2 } | \mathrm{RV}_s} | \mathcal{F}_{s-1}}\\
    &=\re^{zr}\EP{\re^{\left(z \lambda + b - \frac{z^2}{4c} + \gamma z \right) \mathrm{RV}_{s} }\EP{\re^{ c (\epsilon_s - (\gamma - \frac{z}{2c}) \sqrt{\mathrm{RV}_s})^2 } | \mathrm{RV}_s} | \mathcal{F}_{s-1}}\\
    &=\re^{zr - \frac{1}{2} \ln(1-2c) }\EP{\re^{\left( z \lambda + b + \frac{\frac{1}{2}z^2 + \gamma^2c - 2c \gamma z}{1-2c} \right) \mathrm{RV}_{s}} | \mathcal{F}_{s-1}}\,.
  \end{split}
\end{equation}
In the last equality we have used the fact that if $Z\sim\normal$ then 
\begin{equation}
  \label{stnorpro}
  \E{\exp \left( x (Z+y)^2 \right)} = \exp \left( -\frac{1}{2} \ln(1-2x) + \frac{xy^2}{1-2x} \right).
\end{equation}
Using eq.s~(8)-(9) from~\cite{GouJas06} we obtain
\begin{equation}
  \begin{split}
    &\EP{\re^{ z y_{s} + b \mathrm{RV}_{s} + c \scl{\ell}_{s} }| \mathcal{F}_{s-1}}\\  
    &= \exp \left[ zr - \frac{1}{2} \ln(1-2c) - \delta \sclcal{w}( x ,\theta) +  \sclcal{v}( x ,\theta) \left(d + \sum_{i=1}^{p} \beta_i \mathrm{RV}_{s-i} + \sum_{j=1}^{q} \alpha_j \scl{\ell}_{s-j} \right)\right]\,,
  \end{split}
\end{equation}
where 
\begin{equation*}
  \sclcal{v}(x,\theta)=\frac{\theta x}{1 - \theta x}\,, \qquad \sclcal{w}(x,\theta)= \ln \left( 1 - x \theta \right)\,,
\end{equation*}
and
\begin{equation*}
  x(z,b,c) = z \lambda + b + \frac{\frac{1}{2}z^2 + \gamma^2c - 2c \gamma z}{1-2c}\,.
\end{equation*}
From a direct inspection of the relation~(\ref{affine}), we conclude that 
\begin{equation}
  \label{eq:LHARGcal}
  \begin{split}
    \sclcal{A}(z,b,c)   &= zr  - \frac{1}{2} \ln(1-2c) - \delta \sclcal{w}( x ,\theta) + d\sclcal{v}(x,\theta)\,,\\
    \sclcal{B}_i(z,b,c) &=  \sclcal{v}(x,\theta) \beta_i \,,\\
    \sclcal{C}_j(z,b,c) &=  \sclcal{v}(x,\theta) \alpha_j\,. 
  \end{split}
\end{equation} 
Finally, plugging the above expressions for $\sclcal{A}$, $\sclcal{B}_i$ and $\sclcal{C}_j$ in eq.~(\ref{eq:GF-MGFQrecursive}) we readily obtain the recurrence relations under the physical and risk-neutral measures. The no-arbitrage condition similarly follows from formulae~(\ref{eq:LHARGcal}) and relations~(\ref{eq:noARB}) noticing that it is sufficient to impose 
\[
  x(1-\nu_2,-\nu_1,0) = x(-\nu_2,-\nu_1,0).
\]

\section{Risk-neutral dynamics}
\label{app:RNDYNAMICS}
To derive the mapping of the parameters under which the risk-neutral MGF is formally equivalent to the physical MGF, we need to compare eq.~(\ref{coefmgfQ}) to eq.~(\ref{coefmgfP}). In particular we have to find a set of starred parameters for which the recursions under $\mathbb{P}$ correspond to the expressions under $\mathbb{Q}$. More precisely, after defining
\begin{equation*}
  \scl{X}_{s+1}^{**} = z\lambda^* + \scl{B}_{s+1,1}^* + \frac{\frac{1}{2}z^2 + (\gamma^{*})^2 \scl{C}^*_{s+1,1} - 2 \scl{C}^*_{s+1,1} \gamma^* z}{1 - 2 \scl{C}^*_{s+1,1}}\,,
\end{equation*}
the following relations have to hold
\begin{eqnarray}
  \delta   \left( \sclcal{w}( \scl{X}^{*}_{s+1} ,\theta) -  \sclcal{w}(\scl{Y}^*,\theta) \right) &= \delta^* \sclcal{w}( \scl{X}^{**}_{s+1} ,\theta^*)\,,\label{eq:delta}\\
  \beta_i  \left( \sclcal{v}( \scl{X}^{*}_{s+1} ,\theta) -  \sclcal{v}(\scl{Y}^*,\theta) \right) &= \beta_i^* \sclcal{v}( \scl{X}^{**}_{s+1} ,\theta^*)\,,\label{eq:beta}\\
  \alpha_j \left( \sclcal{v}( \scl{X}^{*}_{s+1} ,\theta) -  \sclcal{v}(\scl{Y}^*,\theta) \right) &= \alpha_j^* \sclcal{v}( \scl{X}^{**}_{s+1} ,\theta^*)\,,\label{eq:alpha}\\
  d        \left( \sclcal{v}( \scl{X}^{*}_{s+1} ,\theta) -  \sclcal{v}(\scl{Y}^*,\theta) \right) &= d^* \sclcal{v}( \scl{X}^{**}_{s+1} ,\theta^*)\,,\label{eq:d}
\end{eqnarray}
with $\scl{Y}^* = - \lambda^2/2 - \nu_1 + \frac{1}{8}$.
Eq.~(\ref{eq:delta}) can be rewritten as
\[
  \delta\log\left[1-\frac{\theta}{1-\theta \scl{Y}^*}\left(\scl{X}_{s+1}^*-\scl{Y}^*\right)\right] = \delta^* \log\left(1-\theta^*\scl{X}_{s+1}^{**}\right)\,,
\]
from which we obtain the sufficient conditions $\delta^* = \delta$, $\theta^* = \theta/(1-\theta \scl{Y}^*)$, and $\scl{X}_{s+1}^*-\scl{Y}^*=\scl{X}_{s+1}^{**}$. It is possible to verify by substitution that the latter relation is satisfied posing $\lambda^* = -1/2$ and $\gamma^* = \gamma + \lambda + 1/2$. The relation~(\ref{eq:beta}) is equivalent to
\[
  \frac{\beta_i}{1-\theta \scl{Y}^*} \frac{\theta}{1-\theta \scl{Y}^*}\frac{\scl{X}_{s+1}^*-\scl{Y}^*}{\left[1-\theta/(1-\theta \scl{Y}^*)\left(\scl{X}_{s+1}^*-\scl{Y}^*\right)\right]}=\beta_i^*\frac{\theta^*\scl{X}_{s+1}^{**}}{1-\theta^*\scl{X}_{s+1}^{**}}\,, 
\]
which implies $\beta_i^* = \beta_i/(1-\theta \scl{Y}^*)$. Similar reasoning applies for eq.s~(\ref{eq:alpha}) and (\ref{eq:d}).
\end{document}